# Practical cosmology and cosmological physics

Yu. Baryshev[1], I. Taganov[2], P. Teerikorpi[3]

[1] Astronomical Institute of the St.-Petersburg State University, St.-Petersburg, Russia
[2] Russian Geographical Society, Saint Petersburg, Russia
[3] Tuorla Observatory, Turku University, Turku, Finland

**Abstract:** We present a summary of the International conference "Problems of practical cosmology", held at Russian Geographical Society, 23-27 June 2008, St.-Petersburg, Russia, where original reports were offered for discussion of new developments in modern cosmological physics, including the large scale structure of the Universe, the evolution of galaxies, cosmological effects in the local stellar systems, gravity physics for cosmology, cosmological models, and crucial observational tests of rival world models. The term "Practical Cosmology" was introduced by Allan Sandage in 1995 when he formulated "23 astronomical problems for the next three decades" at the conference on *Key Problems in Astronomy and Astrophysics* held at Canary Islands. Now when the first decade has passed, we can summarise the present situation in cosmological physics emphasizing interesting hot problems that have arisen during the last decade. Full texts of all reports are available at the web-site of the conference.

## 1. What is Practical Cosmology?

Modern cosmology has entered a wonderful epoch – some call it The Golden Age of Cosmological Physics and as a part of physics it should also be a practical science. Indeed, the tremendous growth of observational efforts directly devoted to cosmological questions shows that cosmology is becoming a mature physical science with its own subject and methods. This is a novel and promising situation for a field which up to recent times has been characterized by a respectable collection of theoretical ideas, but a small number of crucial observations to constrain them. Only half a century ago Hermann Bondi (1952) had to express the state of cosmology at the time so that "the checking of a prediction, which usually forms such a vital link in the formulation of physical theories, does not occur in this field, and we have to rely on less objective and certain criteria, such as how satisfying and how simple a theory is".

Practical cosmology is a science on large-scale physics, which deals with world models and experiments planned for testing them. Cosmological scales of distances, times, and masses are the largest ones available for science. Both astronomical observations and fundamental physics are needed for the study of the realm of galaxies – the cosmological laboratory. Practical cosmology has an ambitious goal to build a trustworthy world model, which in itself is a major goal of science and is also a necessary tool for interpreting deep-space phenomena of distant celestial bodies whose distances, sizes and luminosities we otherwise cannot infer.

As all physical sciences, cosmology is based on theory and experiment. Here the theory is the world models, which include cosmological principles and theories of all physical interactions. Experiments consist of astronomical observations (detection of cosmic particles such as photons, protons, neutrinos, gravitons) and the analysis of gathered data.

The advancement of cosmology is determined by the growth of observational data and, on the other hand, by the development of fundamental physical theories. Practical cosmology is the science which makes a link between observation and theory. The major goal of practical cosmology is to develop strategies for uncovering and attacking cosmological problems. Even with the wonderful advanced observational methods available, successful cosmological tests require that we know how to detect and handle different severe selection effects, which may be hidden both in data and, even seemingly secure, methods of data analysis.

In the 20th century Edwin Hubble and Allan Sandage started cosmology as a genuine experimental science, with its own methodology. In his Rhodes Memorial Lectures at Oxford (where he studied in 1910–13), given in 1936 and published as the book *The Observational Approach to Cosmology*, Hubble (1937) stated that: ``*The observable region of space, the*

*region that can be explored with existing instruments, is a sample of the universe. If the sample is fair, its observed characteristics should furnish important information concerning the universe at large."*

These words contain the gist of practical cosmology. ``Our sample of the universe'' is the Local Universe, where observational data are the most accurate. Within about 1000 Mpc this region of space gives the bulk of information on the galaxy universe. This precious region, where reality is in contact with theory, is the starting point for building world models that attempt to extend our cosmic picture far beyond the observable limits.

The observational approach to cosmology foreseen by Hubble was erected on a firm basis by hard work of many astronomers during the 20th century. The beginning of practical cosmology may be related to the classical paper by Sandage (1961) *The ability of the 200-inch telescope to discriminate between selected world models*. Since then cosmology has been a physical, even experimental science with well defined methods to test cosmological models. Sandage formulated what now are called the classical cosmological tests. These include number-magnitude $N(m)$, number-redshift $N(z)$, magnitude-redshift $m(z)$, angular size-redshift $\theta(z)$, surface brightness-redshift $J(z)$, and age-redshift $t(z)$ relations. It was hoped that these predictions could be used with a large telescope and distant objects in order to decide between different Friedmann world models. At that time the first task of cosmology was determination of two fundamental quantities: the Hubble constant $H_0$ and the density $\rho$ of matter.

Gradually, it has been realized that severe problems complicate the use of real astrophysical objects as test bodies for the models. Selection effects and poorly known evolution in look-back time easily hide from view true model parameters and may even deceive the analyst of the observational data into interesting, but erroneous conclusions.

Only on the verge of the 21st century the development of astronomical observing techniques made it possible to detect stellar standard candles at large distances and to make a new step in Sandage' program that he founded in 1961. In 1979 Gustav Tammann, the long-time associate of Sandage, had proposed observations of distant type Ia supernovae as a way to determine whether Einstein's cosmological constant is non-zero. The analysis of the Hubble redshift-distance relation for high-redshift type Ia supernovae led now to the conclusion that the Friedmann model should include an exotic substance, dark energy. "The search for two numbers" in a relatively simple universe containing only dust-like matter and radiation, had to be much extended to include the densities of dark matter and dark energy, and the equation of state of the universe. Currently there are about 15 parameters characterizing the properties of different components in the modern version of the standard cosmological model.

At the conference on Key Problems in Astronomy and Astrophysics held at the Canary Islands, Sandage (1995) presented a list of 23 astronomical problems for the next three decades, in a form analogous to Hilbert's famous 23 problems in mathematics. The following topics in his list are directly related to ``practical cosmology'', the term used by Sandage himself:

- Is the expansion real?
- Evolution in the look-back time
- The distance scale
- Geometry of the universe
- Counts of galaxies
- Nature and amount of dark matter
- Deviations from the pure cosmological expansion
- The intergalactic matter
- Formation time for large scale structure

Since then it has become clear that this list points towards whole directions of cosmological research. Dramatic discoveries have given those problems new significance and deepness.

## 2. From "Precision" cosmology to "Absurd" Universe

During the last decade the cosmic "stage" has become much enriched by new discoveries. Distant supernovae of type Ia, active galaxies and quasars, the thermal background radiation, and very large filamentary structures in galaxy distribution have opened a new page in cosmology. While the possibilities to test cosmological models are improved by the accumulating observational data, they also give rise to new cosmological ideas.

Especially important new results concern the nature and amount of dark matter. These were obtained by Riess et al.(1998) and Perlmutter et al. (1999) who measured supernovae at high redshifts close to one. In fact, these works continued Sandage's program on testing world models by using supernovae of type Ia as standard candles. Within Friedmann models, the observed redshift–magnitude relation for the supernovae requires the addition of the cosmological $\Lambda$ term to Einstein's equations of general relativity. Also observations of anisotropy of the cosmic thermal background radiation by the WMAP satellite (Spergel et al. 2003) confirmed that a dominating dark energy is needed for Friedmann models.

These unexpected findings gave birth to an intensive development of cosmological models containing dark energy and shattered the hope that the universe consists of some ordinary (gravitating) matter having the critical density $\rho_{matter} = \rho_{crit}$, which for long was viewed as a standard model. Now we have arrived at a world where dark energy is the major component and together with dark matter makes the density critical and space Euclidean. Precision cosmology, as now understood, is based on the measurements of anisotropies of the cosmic background radiation by balloon and satellite (e.g. WMAP) experiments. These data can, in principle, give precise values of the main cosmological parameters within the hot big bang scenario if there is no distortion effects caused by our Galaxy and intergalactic medium in the CMBR data.

However this breakthrough into the dark realm has brought into light new puzzling aspects of the modern cosmological model. The last two decades have given us a better understanding of what the real enigmas in cosmology are. At the same time the scope of practical cosmology has grown. The overly optimistic hope that cosmology is solved (Turner 1999) was in a few years turned into the crisis of the cold dark matter model at small scales of galagtic halos (Spergel & Steinhardt 2000; Tasitsiomi 2002; Zackrisson et al. 2006) and at large scales of 100 Mpc structures (Sylos Labini et al. 2008) and the challenge of the unknown physics of the *dark sector* (Peebles & Ratra 2003).

Indeed, our standard cosmological model demands that the dynamics of the whole universe be completely determined by two substances with unknown physics – by the non-baryonic dark matter particles (about 30 percent of the total mass density) and by the negative-pressure substance, dark energy (70 percent of the density).

The dream of "precision cosmology" has shifted to a vision of new cosmological physics (Peebles 2002), meaning that cosmologists now think they know the values of the main cosmological parameters of the standard model, but they do not know the physical sense of the substances which the parameters refer to (Turner 2002). Turner (2003) even characterized the modern state of cosmology as "Absurd universe". Furthermore, several conceptual problems of the standard cosmology recently have been re-discussed (Gron& Elgaroy 2006, Baryshev 2006, Francis et al. 2007). The existence of intriguing paradoxes makes the standard model if not more "absurd", at least more exotic than usually thought.

Understanding the nature of the main cosmological substances, in addition to theoretical considerations, is only possible via cosmological tests – true tools of practical cosmology. For example, due to the remarkable ``flexibility" of the Friedmann model, one may test on a phenomenological level even interaction between dark matter and dark energy (Teerikorpi et al. 2003, Gromov et al. 2004)

## 3. Crucial subjects of cosmological physics discussed at PPC-08 conference

*Session "Large Scale Structure of the Universe"*, chairman **Francesco Sylos Labini** (Enrico Fermi Center and Istituto dei Sistemi Complessi, Rome, Italy), was devoted to the following subjects:
- The history of observational LSS studies
- Statistical physics methods for cosmic structure analysis
- The structure and dynamics of the Local Volume
- Results of analysis of galaxy redshift surveys
- Gravitational lensing and observations of the distribution of dark matter
- Cosmological N-body simulations in comparison with observed structures

Observed properties of the distribution of luminous and dark matter form the basic element of any cosmological model. Therefore a main task of practical cosmology is to develop methods of the large scale structure analysis. One of the most surprising discoveries of 20th century observational cosmology was the complex filamentary structure of the spatial distribution of galaxies. The analysis of this distribution has revealed a new fundamental empirical cosmological law: the power-law behavior of the galaxy correlations.

*The fractal view of the large-scale structure of the Universe.* The last two decades have witnessed the first extensive surveys of galaxy redshifts, which have allowed astronomers to move from the study of the apparent patterns of galaxies on the celestial sphere to the analysis of their three-dimensional distribution in space. The redshift surveys have revealed a rich variety of structures in the galaxy universe, characterized using terms such as binaries, triplets, groups, rich, regular, and irregular clusters, walls, superclusters, voids, filaments, cells, soap bubbles, sponges, great attractors, clumps, concentrations, associations… Of course each form of structure deserves separate studies, but they can be also viewed as natural appearances of one global master entity called fractal. The concept of fractal delivers a convenient mathematical apparatus for a quantitative description of complex stochastic structures with regular long-range power-law correlations.

With fractal techniques one can describe an inhomogeneous galaxy distribution by means of such basic concepts as the fractal dimension, lacunarity, multifractality and others. The fractal dimension of the total mass (luminous and dark) distribution determines the universal mass – radius power law relation, which plays the role of a source of gravity field in cosmological solutions, and hence presents an essential part of cosmological model.

*The debate on fractality.* The discovery of the strongly inhomogeneous spatial distribution of galaxies, on scales from galaxies to superclusters, over four orders of magnitude in scale, is of profound cosmological significance. Only faintly anticipated from photographic surveys, the surprisingly rich texture of galaxies became visible thanks to a large progress in measuring distances by redshifts for thousands of galaxies. The observed clustering is not just random clumping, but exhibits universal long-range power-law correlations.

Remarkably, in the very decade when the galaxy universe was found and theoretical cosmology made it first steps, Selety and Einstein started the debate on cosmological significance of the hierarchical matter distribution. In fact, their correspondence outlined the main directions for future study of nonuniform cosmological models. In observational cosmology the debate on the nature and distances of spiral nebulae went over to a struggle on the structure and extension of the galaxy clustering. This sharp and sometimes dramatic "Debate on Fractality" has been in the limelight almost the whole 20th century and still is. It has involved such persons as Einstein, Hubble, Sandage, Peebles, Charlier, Selety, Lundmark, de Vaucouleurs, Mandelbrot, Pietronero and many others who have studied the large-scale galaxy distribution.

*Power-law correlation of large scale galaxy distribution.* Studies of the large scale distribution of galaxies, which use the largest existing redshift surveys (2dF, the last releases of SDSS), concern the space within about 600 Mpc/h. These may essentially constrain world models. But it is not easy to extract reliable information on the spatial distribution of galaxies from the resulting 3D maps, and one requires appropriate methods of data analysis. Problems related to these methods have created a debate around fractality versus uniformity.

The assumption of the homogeneous distribution of matter in the universe, or Einstein's Cosmological Principle, is a fundamental element in cosmology. Modern observing methods allow a direct check of the spatial distribution of galaxies. It is a critical test of this principle for the luminous matter on large scales. The assumption of a fractal matter distribution, or Mandelbrot's Cosmological Principle, is a generalization of Einstein's homogeneity principle.

*Mechanism of structure formation.* Numerical simulations of gravitating particle systems are the basic way to study the formation of large scale structures within the cold dark matter models. Among the major problems in N-body simulations are the choice of the initial conditions and the representation of the cosmological fluid by a discrete particle set. On small scales, it is a problem how to explain the observed density profiles of different types of galaxies. On large scales, it is still an open question, how the observed (mega)fractality has emerged from the tiny initial fluctuations.

An important consequence of the existence of the hypothetical dark substance is that one cannot yet build a trustable model of the large scale structure formation. In this connection Peebles (2002) makes the illuminating comment that the main unknown element in the standard model is the physics of the dark sector and therefore tests of fundamental physics have a high priority, as they may clear up the nature of the dark substance in the Universe. Until this is done structure formation is a hazardous basis for testing cosmological models.

At the "Large Scale Structure" section the following reports were discussed:

*A Brief History of Large Scale Structures: from the 2D Sky to the 3D Maps.* **P. Teerikorpi**
*Correlations and clustering in the universe.* **F. Sylos Labini**
*Practical cosmology with the Local Volume galaxies.* **I.D. Karachentsev**
*Viewing dark matter with weak gravitational lensing from HST.* **R. Massey**
*Correlations and structures in modern galaxy redshift surveys.* **N.L. Vasilyev**
*Clustering of visible matter and model dark matter halos on different mass and spatial scales.* **A.V. Tikhonov, A.I. Kopylov, S. Gottloeber, G. Yepes**
*A search for super-large structures in deep galaxy surveys.* **N.V. Nabokov, Yu.V. Baryshev**
*Galaxy groups in LCDM simulations and SDSS DR6.* **P. Nurmi, P. Heinämäki, S. Niemi, E. Saar, E. Tago, M. Einasto, E. Tempel, J. Einasto and V.J. Martínez**
*Clustering and Velocities of Quasars from SDSS.* **G. Ivashchenko, V.I. Zhdanov**
*Studying the Ursa Major Supercluster of Galaxies.* **F.G. Kopylova, A.I. Kopylov**
*Properties of Nearby Groups of Galaxies.* **S.-M. Niemi, P. Nurmi, P. Heinämäki, M. Valtonen**
*Environmental effects in the cluster Abell 85 (z=0.055). An HI Imaging Survey and a Dynamical Study.* **H. Bravo-Alfaro, J.H. van Gorkom, C. Caretta**
*Velocity Field in the Local Volume.* **D.I. Makarov**
*Kinematics of galaxy groups: vacuum or fractal acceleration?* **A. Raikov, V. Orlov**
*Autocorrelation Function for Radio Galaxies.* **W. Godłowski**
*Nearby quasars in SDSS.* **P. Heinämäki, P. Nurmi, E. Tago, E. Saar, J. Liivamägi, E. Tempel, M. Einasto, J. Einasto, H. Lietzen, L. Takalo**
*Possible explanation of the Arp-Burbidge paradox.* **B.V. Komberg, S.V. Pilipenko**
*To a question on possible mesh large-scale structure of the universe.* **G.V. Zhizhin**
*The entire-sky catalog of isolated galaxies selected from 2MASS.* **S.N. Mitronova, I.D. Karachentsev, V.E. Karachentseva, O.V. Melnyk**

*Session "Evolution of Galaxies",* chairman **Vladimir P. Reshetnikov** (St.-Petersburg University, St.-Petersburg, Russia), was devoted to the following subjects:
- Observed properties of galaxies
- Dark matter halos – theory and observations
- Spectral energy distribution simulations
- Models of the galaxy evolution

Galaxies are the building blocks of the Universe and an important goal of cosmology is to understand how galaxies form and evolve. In the framework of standard hierarchical models, the formation and evolution of galaxies is a continuous, ongoing process where the observable properties of galaxies are a function of their merging histories, masses and environments.

During the last years, many international projects (SDSS, 2MASS, 2dF, HST deep fields etc.) have been carried out that have distinctively improved the possibilities of modern extragalactic astronomy and cosmology. Observational data on the structure of our Milky Way and of other galaxies have increased by dozens and hundreds of times. For the first time, it has become possible to study the evolution of galaxies and their large-scale structure starting almost from the moment of their formation up to the present epoch.

One of the most actual and promising problems today is to confront observational data obtained for the galaxies at different redshifts with predictions of numerical and semianalytical models. In this way, we can hope to construct the full picture of galaxy formation and evolution during the nearest decades.

At the "Evolution of Galaxies" section the following reports were presented:

*Spiral galaxies at $z \sim 1$.* **V.P. Reshetnikov**
*Stellar Population Modeling of Galaxies in Nearby Groups.* **L.N. Makarova, D.I. Makarov**
*Internal Structure in virialized halos of dark matter.* **E.V. Mikheeva, A.G. Doroshkevich, V.N. Lukash**
*Small-scale dark matter clumps in the Galactic halo.* **V. Berezinsky, V. Dokuchaev, Yu. Eroshenko**
*Hydro-Gravitational-Dynamics Interpretation of the Tadpole VV29 Merging Galaxy System: BDM-Halo Star-Cluster-Wakes.* **C.H. Gibson**
*Cusp Slope Limit Analysis of Double Image Gravitational Lenses.* **P. Mutka**
*Microlensing events in gravitationally lensed quasar Q2237+0305: stars or dark matter.* **A.A. Minakov, R.E. Schild, V.G. Vakulik, G.V. Smirnov, V.S. Tsvetkova**
*Constraining the nature of galaxy haloes with gravitational mesolensing of QSOs by halo substructure objects.* **Yu.L. Bukhmastova, Yu.V. Baryshev**
*Towards the origin theory of SB galaxies containing ring structure.* **S. Nuritdinov**
*Some Thoughts on Dynamical Evolution of Galaxies.* **L.P. Ossipkov**

*Session "The Earth, the Solar System, and Stellar Systems for Cosmology",* chairman **Yurij N. Gnedin** (Central Astronomical Observatory of RAS, Pulkovo, St.-Petersburg, Russia), was devoted to the following subjects:
- Solar System for testing gravity and cosmology
- Dark matter and dark energy in stellar objects
- Influence of the cosmological vacuum on the stellar systems
- The evolution of the Earth

The understanding of the cosmological constant $\Lambda$ is one of the most outstanding topic in modern astronomy and physics. Though the cosmological constant is motivated mainly by observations on very large scales, in principle it may detected on every physical scale. Measuring local effects of $\Lambda$ would be a fundamental confirmation of the detection of dark energy. Some tests of gravity theories, such as the periastron shift, geodesic precession, and the change in mean motion, may be applied to the Solar System in order to constrain the cosmological constant.

Accurate measurements of Earth and Mars the perihelion shifts of the Earth and Mars have provided the tightest bounds on $\Lambda$ among the Solar System tests. The expected secular increase of the Astronomical Unit, as recently reported by Krasincky and Brumberg, allows also one to set the most stringent constraints on the cosmological constant and also to test various gravity theories which explain the observed cosmological acceleration without dark energy. Particularly, the effect of this secular increase of the Astronomical Unit allows us to test the highly popular Dvali-Gabadadze-Porrati multi-dimensional braneworld scenario.

The origin of dark matter is one of the central problems of astronomy and physics. The various pseudoscalar particles are perspective candidates for dark matter. The existence of weakly coupled light pseudoscalar particles may be tested by observations of the solar radiation especially in hard (X and gamma rays) electromagnetic waves. A new idea is to search for primordial quark nuggets among near-Earth asteroids. Primordial quark nuggets have been predicted to contain most of the baryonic number of the Universe. It has been suggested by J.E.Horvath to search for these nuggets in the asteroidal-mass range. Since the strange quark matter is expected to have a plasma frequency as high as 20 Mev, the bare quark surface would act as a perfect mirror to the incident solar light. As a result, one can expect that such a nugget looks like a larger normal asteroid but with an abnormal ratio between the visual and the infrared fluxes and without any emission and absorption lines. Another widely discussed phenomenon that may be related to the local distribution of dark matter is the abnormal additional acceleration of the cosmic Pioneer spacecraft.

General relativistic effects in astrophysical systems have been detected thanks to accurate astrometric observations. We may mention Mercury's perihelion precession, the oblateness of the Solar disk, the relativistic light deflection, the lunar geodetic precession, and the Lense-Thirring precession among others. New astronomical issues that need be discussed include testing general relativity in the gravitational field of the Moon with special cosmic missions and the measurement of the deflection of light from Jupiter. Gravitoelectric and gravitomagnetic effects from the Moon and the satellites of the Solar System can be calculated and compared with the present-day orbit accuracy of modern missions.

At the "Earth, the Solar System, and Stellar Systems for Cosmology" section the following reports were presented:

*Local effects of the Cosmological Constant.* **M. Nowakowski, I. Arraut**
*Effects of cosmology on solar and stellar systems.* **M. Sereno, Ph. Jetzer**
*Relativistic discs in black-hole spacetimes with cosmological constant.* **P. Slany, Z. Stuchlik**
*Dark Matter in the Solar System.* **V.L. Kauts**
*On the cosmological nature of Pioneers anomalous acceleration.* **A. Raikov, V. Orlov**
*Planets and dark energy.* **C. Gibson, R. Schild**
*Geothermal heat flow as a problem: the history and the present state.* **A. Zemtsov**
*Terrestrial Origin Versus Extraterrestrial Overflow of Thermodynamics.* **G. Guzzetta**
*Detection and Study of Dark Electric Matter Objects – presumably Planckian black holes.*
    **E.M. Drobyshevski, M.E. Drobyshevski**
*Influence of the Cosmological Lambda-Term on the Parameters of the Earth–Moon System.*
    **Yu.V. Dumin**
*Observational Evidences of the Cosmological Deceleration of Time.* **I.N. Taganov**

*Session "Gravitation Physics for Cosmology",* chairman **Theo Nieuwenhuizen** (Institute for Theoretical Physics, Amsterdam, The Netherlands ), was devoted to the following subjects:
- Problem of the energy-momentum for the gravitational field
- General relativity and testing gravitation physics
- Alternative gravitation theories which allow crucial tests
- Compact relativistic objects for testing gravitation theories
- Gravitational waves as test of gravitation theories

Any cosmological model is a particular solution of the gravitational field equations for the matter filling the Universe. Gravitation is that universal force which acts between all matter and rules the dynamics of the whole Universe.

General relativity is the basic theory of the standard cosmology. However, it is not a quantum theory and recent developments in theoretical physics suggest that other possibilities exist to construct a theory that could replace Einstein's general relativity. Thus it is natural for practical cosmology to study rival theories.

As gravitation is the true phenomenon at large scales, cosmology is also an important test-bench for gravitation theories. It is important to develop alternative ideas about gravitation up to the level where rival theories give different predictions, leading to crucial observational tests. Usually, within a theory there are some undetermined parameters that may help one to solve arising problems, until a new powerful test appears. Only a lengthy process of testing alternative theoretical ideas will finally lead to a theory which explains the empirical facts by the smallest number of additional parameters.

The problem of gravitation quantization is linked to the nature of gravitational interaction. Indeed, if gravitation is geometrical in nature (a property of curved space), then one should develop methods of space-time quantization. But if gravitation is a force mediated by gravitons (quanta of a relativistic field), then one should find methods and new principles for overcoming the non-renormalizability. Space-time foam then should not exist. From the Practical Cosmology point of view it is important to discuss future astrophysical observations and space experiments which may distinguish between alternative approaches to the physical description of the gravitational interaction.

Nowadays the gravitation physics and cosmology are so closely related that a proper gravitation theory may solve simultaneously the problems of dark matter and dark energy in cosmology and the problem of the existence of horizons in locally observed astrophysical objects.

At the "Gravitation Physics for Cosmology" section the following reports were presented :

*Energy-momentum of the gravitational field: crucial point for gravitation physics and cosmology.* **Yu. Baryshev**
*SN1987A revisited.* **G. Pizzella**
*Gravitational collapse as the source of gamma-ray bursts.* **V. Sokolov**
*Cosmological Properties of Eternally Collapsing Objects (ECOs).* **A. Mitra**
*Why Are Some Quasars Radio Loud?* **R. Schild**
*On the relativistic theory of gravitation and exact solutions for the interior of black holes.* **Th.M. Nieuwenhuizen**
*Cosmological solution of the relativistic theory of gravitation and supernova observational data.* **A.V. Genk, A.A.Tron**
*Gauge theory of gravitation: electro-gravity mixing.* **E. Sanchez-Sastre, V. Aldaya**
*Gravitational Thermodynamics, the Cosmological Constant and Quantum Gravity.* **F.A. Assaleh**
*New possibilities for observational distinction between geometrical and field gravity theories.* **Yu. Baryshev**

*Session  "Cosmological Models and Crucial Observational Tests",* chairman **Yurij V. Baryshev** (Astronomical Institute of St.-Petersburg University, St.-Petersburg, Russia) , contained two parts devoted to theoretical models and their observational tests.

*Constructive cosmology.* Practical cosmology has the goal to construct the true model of the *real* Universe, and this constructive side is paralleled by its exciting explorative aspect, in its penetration into deep space. Due to limited observational means and theoretical understanding at each epoch, the adopted world model has its limits, too, even if it may seem quite satisfying. Indeed, one might be content with "fine-tuning" the current model. However, in cosmology it is advisable to probe different ways to explain observations and alternative initial hypotheses. This also leads one to classify reasonable cosmological models. Highlighting their cornerstones, it may be helpful for planning crucial tests and even directing the thinker to a novel idea.

In world models, old and new alike, one may discern three cornerstones. All start with some *Observation*. All use some *Theory*. All rely on some *Cosmological Principle*. We may characterize cosmological models by asking about their relation to those fundamental things.

- What observations are viewed as cosmologically important or relevant?
- What physical theories, in particular for gravitation, are used?
- What fundamental assumptions, going beyond our finite empirical range, are expressed as Cosmological Principle(s)?

Our present standard (Einstein-Friedmann) world model is based on key facts in the large scale realm of galaxies, where we see the cosmological redshift and the Hubble law, and the cosmic background radiation, very isotropic and accurately thermal. Einstein's general relativity and Friedmann's expanding model form the theoretical framework, together with the standard particle physics. Friedmann's model rests on Einstein's Cosmological Principle of homogeneous and isotropic spatial distribution of all main matter components.

Thus we see that world models may be grouped according to a few key questions. One set concerns the framework itself: What is gravity? What are the matter components and their equations of state? How are these components distributed in space? The second set touches observations and inferred cosmological laws: What causes the cosmological redshift? What is the origin of the thermal background radiation? What is behind the global evolution and the arrow of time? What is the spatial and temporal extent of the Universe?

*Gravity theory.* The heart of any cosmological model is the gravity theory, as gravitation is the dominant force beyond the scale of stars. The gravity theory can be constructed by different ways, with at least two main understanding of relativistic quantum gravity -  as a geometry or as a material field . Within the geometric approach there are attempts to generalize Einstein's gravity theory in order to avoid dark matter and dark energy.

It is important to note that up to now relativistic gravity has been tested experimentally only in weak field conditions. The well known classical  relativistic effects usually cited in favor of general relativity may be derived within the field gravity theory, too. It is true that expanding space (and the Hubble law) follow elegantly from the general relativity-based Friedmann model. However, for the very reason that the redshift is a primary cosmological phenomenon, independent tests of its physical origin should be done. To prove that space is really expanding, would mean strong support for general relativity as the gravity theory, and vice versa the absence of space expansion, i.e. the other physical mechanism for cosmological redshift, will support the field gravity theory.

Although geometry has had great success in gravity physics, there are conceptual difficulties within such a description of gravity, including the problem of the pseudo-tensor character of the energy-momentum tensor for the gravity field. These may signal a need for a gravity theory where the energy of the gravity field has the same regular sense as in all other fundamental interactions. Developments in theoretical physics suggest that we may be close to a transition from Einstein's general relativity to a quantum relativistic gravity theory, which inevitably will change the cosmological model. Such cosmic constituents like the vacuum and dark energy have a quantum nature. At present they enter general relativity on a phenomenological level only. As parts of the new quantum gravity theory these entities will likely affect our view of such things as dark matter, large-scale structure formation, and other phenomena, including the nature of the cosmological redshift.

*Composition and distribution of matter.* Two main components of cosmic matter are relativistic and non-relativistic substances. The relativistic parts, such as photons, neutrino, gravitons, the cosmological vacuum, have a natural uniform spatial distribution. The non-relativistic component is the matter related to galaxies, containing ordinary luminous matter (stars, gas) and possible dark substances (baryonic and non-baryonic). Modern redshift-based maps and the large-scale structure analysis have led to the result that a stochastic fractal distribution with the dimension $D = 2$ may approximate reality (luminous matter) on scales from 0.1 Mpc up to about 100 Mpc. The distribution and constitution of dark matter is still an open issue.

One should have in mind the possibility that both Einstein's Cosmological Principle of homogeneity and Mandelbrot's Cosmological Principle of fractality could be simultaneously true, but relate to different spatial scales and different components of matter.

*The nature of redshift and the Hubble law.* The cosmological redshift is an observational fact that can be interpreted in various ways. At least three mechanisms exists producing redshifts independent of the wavelength: space expansion, Doppler effect and global gravitational effect. An important constraint for the possible mechanism is given by the linear distance–redshift relation. Harrison (1993) has shown a clear distinction between different interpretations of the cosmological redshift. Gron & Elgaroy (2006) and Francis et al.(2007) have argued directly from the distance–redshift relation that it cannot be due to the ordinary Doppler effect as motion within space. The space expansion governed by uniform matter produces both the redshift and the linear Hubble law. Intriguingly, the global gravitational effect within a fractal non-uniform matter distribution also produces the linear redshift Hubble law without space expansion, so this alternative should also be studied (Baryshev 2006).

*Space expansion.* The scale factor $S(t)$ describes mathematically the space expansion of a universe. All distances between uniformly scattered particles are changing with time as $r(t) = S(t) \times \chi$. As the observed Universe is, after all, highly inhomogeneous, one is compelled to ask what expands and what doesn't. Physically, a way to view the expansion of space is as creation of space together with the physical vacuum. This opens interesting conceptual problems in the physics of space expansion, a subtle subject deserving attention.

*The origin of the CBR.* There are two basically different approaches to the nature of the cosmic background radiation. The first makes use of the standard hypothesis that it originated in the hot early Universe. However, as was shown by Harrison (2000, 1995), the process of cooling of photon's gas in expanding space is based on the violation of energy conservation.

The second approach tries to understand CBR as the result of integrated contributions from radiation sources at a variety of redshifts. E.g., Hoyle (1982, 1991) pointed out that the energy from nuclear reactions and radiated by stars during their life is just the same as the energy of the observed CBR, and if there is a process to thermalize this energy then the background radiation could have this ordinary physical black body origin.

*Alternative frameworks in cosmology.* Amidst the success story of modern cosmology one should not lose sight of a few healthy reminders of why also alternative models have the right to exist in contemporary cosmology.

*First*, the finite observable part of the possibly infinite Universe does not allow one to test directly the initial hypotheses on the Universe as a whole. The possibility of a major reform is never excluded.

*Second*, even the observed key phenomena may have different interpretations, each corresponding to a choice of the basic framework that is able to explain the main cosmologically relevant observations.

*Third*, theoretical physics is a developing subject and "new physics" may offer a wide spectrum of different cosmological applications. Even in the current standard model the nature and physics of 95 percent of the substance is unknown (dark matter, dark energy).

It is also good to realize that alternative approaches (sometimes described as a "noisy minority") actually define that territory of theoretical ideas that push astronomers to devise important cosmological tests. Hubble & Tolman (1935) suggested the number counts and the surface brightness as ways to test alternative causes of the cosmological redshift. Hoyle (1959) proposed the angular size – redshift relation to test the steady state model. Sandage, Tamman & Hardy (1972) viewed the number counts and the linear Hubble law as a test of de Vaucouleurs's (1971) hierarchical model. Novel ideas may also bring into light weak spots of the standard model and may help to find ways to test the underlying cosmological physics.

*Selection effects in astronomy and cosmology.* A modern cosmologist has at his disposal tremendous amounts of data obtained at different wavelength bands. One might think that the more there are objects, the easier it is to test world models. Unfortunately this not exactly so, because collected data are influenced by various selection, distortion, and evolution effects. The observations in the "cosmological laboratory" are inevitably affected by our one-point position
in space-time and all sorts of instrumental limitations. The selection effects distort the original physical relations between observed quantities and are dangerous in that they may make observed relations imitate theoretical dependencies, which are not true and just originate from the observing procedure. A classical example is the notorious Malmquist bias. As one probes deep space, one progressively observes objects which are either apparently fainter or intrinsically more luminous. Because one cannot measure arbitrarily faint fluxes, one necessarily observes exceptionally powerful objects. Such are not representative of the typical populations at such large distances. This "iceberg effect" is one aspect of a group of selection effects often collectively put under the name Malmquist bias.

*The Part I section "Cosmological Models"*, chairman **Igor N. Taganov** (Russian Geographical Society, St.-Petersburg, Russia), was devoted to the following subjects:
- Basic principles and paradigms in cosmology
- Standard cosmological model and alternatives
- Predictions for crucial observational tests

Rapid progress of astronomical instruments and computer technologies in the last quarter of the bygone 20$^{th}$ century enriched cosmology with numerous observational data on the structure of distant cosmos. In spite of it, the amount of quantitative parameters of the entire integral universe grows slowly. Now we may discuss only four Key Cosmological Parameters, which can be estimated by different independent observational methods (the Table):
- Hubble parameter and average universe mass density (from 1930s)
- Energy density and temperature of CMBR - cosmic microwave background radiation (from 1970s)
- Fractal dimension of the universe large-scale structures (from 1980s)

*Table. Key Cosmological Parameters*

| Key Cosmological Parameters | Observations | Quantum Cosmology Estimations (I. Taganov) |
|---|---|---|
| Hubble parameter | $H = (1.6 \div 2.6) \cdot 10^{-18}$ s$^{-1}$ $= (50 \div 80)$ km/s/Mpc | $H = 9G\hbar/16c^2 r_e^3 = 1.970 \cdot 10^{-18}$ s$^{-1}$ = 61.6 km/s/Mpc |
| Average mass density | $\rho_m = (5 \div 12) \cdot 10^{-30}$ g cm$^{-3}$ | $\rho_m = 4H^2/9\pi G = 9\hbar^2 G/64\pi c^4 r_e^6 = 8.217 \cdot 10^{-30}$ g cm$^{-3}$ |
| CMBR energy density and temperature | $\rho_{CMBR} = 4.19 \cdot 10^{-13}$ erg cm$^{-3}$ $T_{CMBR} = 2.728 \pm 0.004$ K | $\rho_{CMBR} = \rho_m e^4/\hbar^2 = 3.929 \cdot 10^{-13}$ erg cm$^{-3}$ $T_{CMBR} = (\rho_{CMBR}/\sigma)^{1/4} = 2.684$ K |
| Fractal dimension of the universe large-scale structures | $\rho_m \propto r^{-1}$ $D = 2 \pm 0.1$ | $\rho_m = (3\hbar/8\pi c r_e^3) r^{-1} = 1.878 \cdot 10^{-1} \cdot r^{-1}$ g cm$^{-3}$ $D = 2$ |

Gravitational constant $G = 6.673 \cdot 10^{-8}$ cm$^3$ g$^{-1}$ s$^{-2}$; Planck constant $\hbar = h/2\pi = 1.055 \cdot 10^{-27}$ erg s; speed of light in a vacuum $c = 2.998 \cdot 10^{10}$ cm s$^{-1}$; charge of electron $e$ ($e^2 = 2.307 \cdot 10^{-19}$ g cm$^3$ s$^{-2}$); classic electron radius $r_e = e^2/m_e c^2 = 2.818 \cdot 10^{-13}$ cm; Stefan-Boltzmann constant $\sigma = 7.566 \cdot 10^{-15}$ erg cm$^{-3}$ K$^{-4}$.

Besides numerous successful qualitative predictions and elegant mathematical analyses, cosmology based on Einstein-Friedmann equations exposed strange Paradox of Theoretical Uncertainty: the absence of theoretical estimations of the Key Cosmological Parameters even with symbolic accuracy. Many alternative models existing in contemporary cosmology reveal the same paradox. Probably in addition to observational tests of newborn cosmological models, it would be reasonable to use at first stage the theoretical test of overcoming the Paradox of Theoretical Uncertainty.

An example of successful defeat of the Paradox of Theoretical Uncertainty demonstrates Quantum Cosmology, which was born long before triumphal evolution of the Standard Cosmological Model. Quantum cosmology could be said to have begun with Max Plank' proposal in the conclusion of his legendary presentation in Academy of Sciences in Berlin on May 18, 1899 to introduce the "natural units" of measurement, basing on his new quantum constant. Plank' idea, however, got no support from his contemporaries, and it was buried in oblivion for more than half a century until in the 1950s John Wheeler rediscovered Planck' fundamental length in his "geometro-dynamics". In 1958 Nikolai Kozyrev achieved an important heuristic result introducing first global cosmological quantum parameter - the "course of time constant" $e^2/\hbar$, but like Planck he had not many followers. Despite occasional criticism, cosmology continued to use Newton-Einstein gravitation theory, abandoning for a long time an idea of the search for specific relativistic and quantum laws of mega-world. This was by no means because the failure to realize limited prospects of a mega-world theory based on Newton-Einstein gravitational equations and thermodynamics. The quest for specific quantum mega-world laws was inhibited, until the last quarter of the 20$^{th}$ century, by inferior, compared to quantum physics, amount of reliable quantitative data from observations of distant cosmic structures.

An important stimulus for progress in Quantum Cosmology was the discovery of fractal geometry of the universe large-scale structures. It appeared that fractal dimension of the universe large-scale structures $D = 2$ is the same as the dimension of a fractal micro-particle trajectory described by quantum mechanics. This heuristic analogy helped to estimate theoretically all the Key Cosmological Parameters (the Table).

At the "Cosmological Models" section the following reports were presented :

*The Standard Cosmological Model.* **V. Lukash**
*Physics of space expansion: the root of conceptual problems of the standard cosmological model.* **Yu.Baryshev**
*Questioning the Observational Evidence for the Cosmological Standard Model.* **A. Unzicker**
*A New Case for an Eternally Old Infinite Universe.* **A. Mitra**
*The Cosmic Defect theory tested by observation.* **A. Tartaglia, M. Capone, N. Radicella**
*Field fractal cosmological model as an example of practical cosmology approach.* **Yu. Baryshev**
*Conception of Quantum Cosmology.* **I.N. Taganov**
*Non-positive Dimension Spaces.* **Yu. I. Babenko**
*Anthropic Cosmological Principle and Universal Cosmological Principle as the basis of theoretical and practical cosmology.* **G. M. Idlis**
*What Happened if Dirac, Sciama and Dicke had Talked to Each Other about Cosmology?* **A. Unzicker**
*The Problem of Observation and Regional Ontologies.* **I.D. Nevvazhay**
*Inertial Frame Transformation Based on Lobachevsky Function and Some Optical Phenomena (Michelson-Morley experiment and Doppler Effect).* **N.G. Fadeev**
*On a possible quantum contribution to the redshift.* **K. Urbanowski**
*Cosmological fractal negative acceleration.* **A.Raikov**
*How to verify the redshift mechanism of low-energy quantum gravity.* **M.A. Ivanov**
*Redshift as a characteristic of speed of light from space objects.* **S. Tiguntsov**
*Alternative cosmological model.* **V.V. Kossarev**
*Universe driven by the vacuum of scalar field: VFD model.* **S.L. Cherkas, V.L. Kalashnikov**
*Off-site continuums as Cosmological Models of Galaxies and the Universe.* **A.V.Novikov-Borodin**
*The new method of Friedmann equations solving and space-time without events horizons.* **A.V. Yurov , V.A. Astashenok**

*The Part II section "Crucial Observational Tests",* chairman **Pekka Teerikorpi** (Tuorla Observatory, Turku University, Turku, Finland), was devoted to the following subjects:
- Testing cosmology using the Sunyaev-Zeldovich effect
- The value of the Hubble constant
- Spectra of quasars in crucial cosmological tests
- Cosmic background radiation and its distortion by foregrounds
- Variability of the fundamental constants

Modern cosmology is an experimental science where observations play role of a test-bench for the theoretical models of the Universe. Here an experiment usually means a carefully planned set of observations directed to test some theoretical predictions. However many distortion effects and selection biases are ready to deceive cosmologist if he does not aware about these practical things always attached to observed relations in astronomy. Physics of 21 century considers the whole observable Universe as a gigantic laboratory, where the physical laws may be investigated at the largest available scales. The cosmological laboratory has many idiosyncrasies that complicate the work of a cosmologist, including non-locality of the observations and many distortion-selection effects always putting their finger on observed relations in astronomy.

The main goal of practical cosmology is to distinguish between different cosmological theories in order to find the best world model, which successfully describes available observational data and is able to predict some new testable phenomena. One may discern two kinds of cosmological tests corresponding to the two parts of any cosmological theory, namely, the initial hypotheses and the cosmological parameters of the model. Those we call crucial and parametric cosmological tests. Crucial tests allow us to judge the validity of the basic assumptions of the theories, and parametric ones are used for experimental estimation of the model parameters.

We view as crucial those tests that concern the validity of the decisively important initial hypotheses of cosmological models or their fundamental predictions:
- testing the validity of physical laws at cosmological scales
- determination of the matter distribution in space
- testing of the reality of space expansion
- measuring the temperature of the CMBR at high redshift
- determination of the ages of the oldest objects
- measuring the evolution of the chemical composition of matter at high redshifts.

Crucial cosmological tests are the most important and the most difficult to perform because they require the highest technological achievements of the time. A cosmological model should give definite mathematical descriptions for observable quantities and predict relations between them. Alternative ideas, which allow experimental verification, play a positive role in cosmology and questions about their feasibility must be answered by the means of suitable observational tests.

At the "Crucial Observational Tests" section the following reports were presented :

*Probing cosmology and astro-particle physics with the SZ effect.* **S. Colafrancesco**
*Opik's method, Eddington's luminosity and Hubble's constant.* **P. Teerikorpi**
*Some difficulties for measuring and interpreting the expansion of the universe.* **G. Paturel**
*Cosmology with objects from the Hamburg qso survey.* **D. Reimers**
*QSO spectra in cosmological tests.* **S.A. Levshakov, I.I. Agafonova and P. Molaro**
*Sensitivity of microwave molecular transitions and atomic FIR transitions from quasar spectra to variation of fundamental constants.* **M.G. Kozlov, V.V. Flambaum, S.A. Levshakov, D. Reimers, S.G. Porsev and P. Molaro**
*CMB data analysis: methods and problems.* **O. Verkhodanov**
*The complexity of dust foreground emission in Cosmic Microwave Background maps.* **X. Dupac**
*High-Accuracy Method for the Removal of Point Sources from Maps of the Cosmic Microwave Background.* **A.T. Bajkova**
*Gamma-Ray Bursts and Practical Cosmology.* **A.S. Moskvitin, E. Sonbas, I.V. Sokolov, T.A. Fatkhullin**
*Properties of WMAP cross-sections in the field of the RATAN-600 survey.* **M.L. Khabibullina, O. Verkhodanov, Yu. Parijskij**
*Giant radio galaxies: problems of understanding and problems for CMB?* **O.V. Verkhodanov, M.L. Khabibullina, M. Singh, A. Pirya, N.V. Verkhodanova, S. Nandi**
*Ultra-steep spectrum decametric sources for cosmological researches.* **O.V. Verkhodanov, N.V. Verkhodanova, Heinz Andernach**
*New Method of Data Mining in Practical Cosmology.* **V.V. Vitkovskiy, V.L. Gorokhov**